\documentclass[%
 reprint,
 showpacs,
 amsmath,amssymb,
 aps,
]{revtex4-1}
\usepackage{amsmath}
\usepackage{cases}
\usepackage{amssymb}
\usepackage{CJK}
\usepackage{graphicx}
\usepackage{dcolumn}
\usepackage{bm}
\usepackage{color}

\begin{document}
\begin{CJK*}{GBK}{} 

\preprint{APS/123-QED}

\title{Isobaric Yield Ratio Difference in Heavy-ion Collisions, and Comparison to Isoscaling}

\author{Chun-Wang MA
$^{1}$}\thanks{Email: machunwang@126.com}
\author{Shan-Shan WANG
$^{1}$}
\author{Yan-Li ZHANG
$^{1}$}\author{Hui-Ling WEI
$^{1}$}

\affiliation{$^{1}$ Department of Physics, Henan Normal University, \textit{Xinxiang 453007}, China
}




\date{\today}

\begin{abstract}
An isobaric yield ratio difference (IBD) method is proposed to study the ratio of the difference
between the chemical potential of neutron and proton to temperature ($\Delta\mu/T$) in heavy-ion
collisions. The $\Delta\mu/T$ determined by the IBD method (IB-$\Delta\mu/T$) is
compared to the results of the isoscaling method (IS-$\Delta\mu/T$),
which uses the isotopic or the isotonic yield ratio. Similar distributions of
the IB- and IS-$\Delta\mu/T$ are found in the measured 140$A$ MeV $^{40,48}$Ca + $^{9}$Be
and the $^{58,64}$Ni + $^{9}$Be reactions. The IB- and IS-$\Delta\mu/T$ both have a distribution with
a plateau in the small mass fragments plus an increasing part in the fragments of relatively larger
mass. The IB- and IS-$\Delta\mu/T$ plateaus show dependence on the $n/p$ ratio of the projectile.
It is suggested that the height of the plateau is decided by the difference between
the neutron density ($\rho_n$) and the proton density ($\rho_p$) distributions of
the projectiles, and the width shows the overlapping volume of the projectiles in
which $\rho_n$ and $\rho_p$ change very little. The difference between the IB-
and IS-$\Delta\mu/T$ is explained by the isoscaling parameters being constrained by
the many isotopes and isotones, while the IBD method only uses the yields of two isobars.
It is suggested that the IB-$\Delta\mu/T$ is more reasonable than the IS-$\Delta\mu/T$,
especially when the isotopic or isotonic ratio disobeys the isoscaling. As to the question
whether the $\Delta\mu/T$ depends on the density or the temperature, the density dependence
is preferred since the low density can result in low temperature in the peripheral reactions.
\end{abstract}

\maketitle
\end{CJK*}


\section{introduction}

In the models based on the free energy to predict the fragments in heavy-ion collisions
(HICs) above the Fermi energy, the yield is mainly determined by the free energy, the
chemical potential of proton and neutron, the temperature, etc. \cite{AlbNCA85DRT,Bot02-iso-T,Isoscaling,HShanPRL}.
In the ratios between the fragment yields, some of the information which the fragment carries
will cancel out, and the retained information is useful to study the properties of the
colliding sources \cite{Isoscaling,MBTsPRL01iso}, the fragment itself \cite{ModelFisher3,Huang10},
and the temperature of the reactions \cite{AlbNCA85DRT,MaCW12PRCT}. The isoscaling
method is one of the important methods to constrain the symmetry energy of the nuclear
matter in HICs \cite{IS-fluctuation13,Bot02-iso-T,Isoscaling}, which makes it important
for the study of the nuclear equation of state \cite{BALi08PR}. The isoscaling phenomena are
systemically studied experimentally, and extensively examined in theories from dynamical
models to statistical models \cite{ChenZQ10-iso-sym,Huang10NPA-Mscaling,Ono03RRCiso,Ono04RRCiso,MBTsPRL01iso,SouzaPRC09isot,Soul03-iso,Soul06-iso-T-sym,TianWD05-iso-CPL,Dorso06-iso-finite,Fang07-iso-JPG,FuY09isoCaNi},
The effects of the secondary decay, which significantly influence the results, are also
investigated \cite{ZhouPei11T,Tsang01-iso-sym,Liu04-iso-sym,Tsang06-iso,FuY09isoCaNi}.
Besides the isoscaling method, the isobaric yield ratio is promoted to determine the
symmetry energy of the fragments produced in HICs in a modified Fisher model
\cite{Huang10, MA12CPL09IYRAsbsbv,MaCW11PRC06IYR,MaCW12EPJA,MaCW12CPL06}. At the same
time, the isotopic ratio and the isobaric ratio are also used to study the temperatures
of the colliding sources \cite{AlbNCA85DRT,Goodman84Tc,YgMa99PRCTc,YgMa05PRCTc,Nato95T,Wada97T,SuJun11PRCTiso,JSWangT05,TrautTHFrg07}
or the heavy fragments in HICs \cite{MaCW12PRCT,TrautTHFrg07,MaCW13ComTheoPhys}. Since both the isoscaling method and the isobaric ratio method are deduced in the framework
of the free energy theories, and they both relate the yield of fragments to the symmetry
energies of the colliding sources, it is important to compare the nuclear symmetry energy
determined by them.

In this article, the difference between the chemical potentials of the neutron and proton
will be compared using the isoscaling method and the isobaric ratio method. In Sec. \ref{theory},
the isoscaling and the isobaric ratio difference methods will be deduced in the framework
of the grand-canonical ensembles. In Sec. \ref{disc},  the fragment yields in the
140$A$ MeV $^{40,48}$Ca + $^9$Be and $^{58,64}$Ni + $^9$Be reactions will be analyzed
using the isoscaling method and the isobaric yield ratio (IYR) method, and the results will be compared. A
summary will be presented in Sec. \ref{summary}.

\section{model description}
\label{theory}
The isoscaling and the isobaric ratio difference (IBD) methods in the grand-canonical
ensembles will be introduced briefly. In the grand-canonical limit, the yield of
a fragment with mass $A$ and neutron-excess $I$ ($I = N-Z$) is given by \cite{GrandCan,Tsang07BET}
\begin{equation}\label{yieldGC}
Y(A,I) = CA^{\tau}exp\{[F(A,I)+\mu_{n}N+\mu_{p}Z]/T\},
\end{equation}
where $C$ is a constant. $N$ and $Z$ are the neutron and proton numbers.
$\tau$ is nonuniform in different reaction systems \cite{Huang10Powerlaw}. $\mu_n$
and $\mu_p$ are chemical potentials of the neutron and proton, respectively; $F(A,I)$
is the free energy of the cluster (fragment), and $T$ is the temperature.

The isoscaling method is as follows: for one fragment in two reactions of the
same measurements, based on Eq. (\ref{yieldGC}), the yield ratio of the two
reactions, $R_{21}^{IS}(N,Z)$, can be defined as \cite{Isoscaling,HShanPRL}
\begin{equation}\label{isoscaling}
R_{21}^{IS}(N,Z)=Y_{2}(N,Z)/Y_{1}(N,Z)=C'\mbox{exp}(\alpha N+ \beta Z),
\end{equation}
where $C'$ is an overall normalization constant which originates from the different
reaction systems. $\mu_n$ and $\mu_p$ are assumed to change very slowly;
$\alpha = \Delta\mu_n/T$ with $\Delta \mu_n = \mu_{n2} - \mu_{n1}$, and
$\beta = \Delta\mu_p/T$ with $\Delta\mu_p = \mu_{p2} - \mu_{p1}$, which reflect
the properties of the colliding sources. In the isotopic ratios, $\beta$ cancels out
and $\alpha$ can be fitted; and in the isotonic ratios, $\alpha$ cancels out and
$\beta$ can be fitted. $\alpha \approx -\beta$ is found, and $\alpha$ can be related
to the symmetry energy ($C_{sym}$) in nuclear mass of the colliding source by
$\alpha=4\frac{C_{sym}}{T}[(\frac{Z_1}{A_1})^2-(\frac{Z_2}{A_2})^2]$, or some
similar relationships \cite{SouzaPRC09isot,ChenZQ10-iso-sym,IS-fluctuation13,Ono03RRCiso}.

The isobaric yield ratio (IYR) is used to study the symmetry energy of the fragment
at finite temperatures \cite{Huang10,MaCW11PRC06IYR,MaCW12EPJA,MaCW12CPL06,MaCW12PRCT}.
When using IYR to study the $\mu_n$ and $\mu_p$, the analysis method should be reconstructed.
Starting from Eq. (\ref{yieldGC}), in one single reaction, the IYR between the
isobars differing by 2 units in $I$, $R^{IB}(I+2,I,A)$, can be defined as
\begin{eqnarray}\label{ratiodef}
&R^{IB}(I+2,I,A)
=Y(A,I+2)/Y(A,I)  \nonumber\\
& =\mbox{exp}\{[F(I+2,A)-F(I,A)+\mu_n-\mu_p]/T\},
\end{eqnarray}
The $CA{^\tau}$ term in Eq. (\ref{yieldGC}) cancels out and the system dependence is
removed. Assuming that the isobars in the ratio have the same temperature, only the
retained $\mu_n$ and $\mu_p$ are related to the colliding sources. Taking the logarithm
of Eq. (\ref{ratiodef}), one can obtain,
\begin{equation}\label{tlnRcal}
\mbox{ln}R^{IB}(I+2,I,A)=(\Delta F+\Delta\mu)/T,
\end{equation}
where $\Delta F = F(I+2,A)-F(I,A)$, and $\Delta\mu=\mu_n-\mu_p$. In two reactions of
the same measurements, the difference between the IYRs, i.e. the IBD method, can be defined as,
\begin{eqnarray}\label{DBratio}
\Delta \mbox{ln}R_{21}^{IB}&=\mbox{ln}[R_{2}^{IB}(I+2,I,A)]-\mbox{ln}[R_{1}^{IB}(I+2,I,A)]   \nonumber\\
             & =\Delta\mu_{n}/T-\Delta\mu_{p}/T=\Delta\mu/T=\alpha-\beta, \hspace{0.4cm}
\end{eqnarray}
Eq. (\ref{DBratio}) also shows the relationship between the results of the
isoscaling parameters ($\alpha$ and $\beta$) and the IBD method. For convenience,
the IBD and isoscaling $\Delta\mu/T$ are labeled as IB-$\Delta\mu/T(\equiv\Delta\mbox{ln}R_{21}^{IB})$
and IS-$\Delta\mu/T(\equiv\alpha-\beta)$, respectively.

\begin{figure}[htbp]
\includegraphics
[width=8.cm]{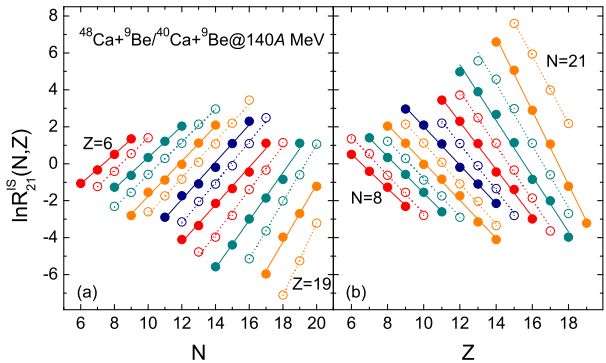}
\caption{\label{ISCa4048} (Color online) (a) The isotopic yield ratio of
the isotopes from $Z$ = 6 to 19, and (b) the isotonic yield ratio of
the isotones from $N$ = 8 to 21 in the 140$A$ MeV $^{40,48}$Ca + $^9$Be
reactions \cite{Mocko06}. The lines are the linear fitting results.
}
\end{figure}

\section{results and discussions}
\label{disc}
The yields of the fragments produced in the 140$A$ MeV $^{40,48}$Ca + $^9$Be and the
$^{58,64}$Ni + $^9$Be reactions were measured by Mocko \textit{et al.} at the
National Superconducting Cyclotron Laboratory (NSCL) in Michigan State University.
The details of the measurements were described in Ref. \cite{Mocko06}. The isoscaling
phenomena in these reactions were studied in Ref. \cite{FuY09isoCaNi}, and the
isotopic (isotonic) yield distributions in these reactions were studied in
Ref. \cite{MaCW09PRC}. In this article, the IB- and the IS-$\Delta\mu/T$
associated with the fragments in these reactions will be analyzed. The analysis
will be performed between the isotopic $^{48}$Ca/$^{40}$Ca and $^{64}$Ni/$^{58}$Ni
reactions, the $n/p$ symmetric $^{58}$Ni/$^{40}$Ca reactions, and
the neutron-rich $^{48}$Ca/$^{64}$Ni reactions. The reaction of
the relatively small $n/p$ projectile is denoted as 1, and the other one as 2.
The isoscaling parameters $\alpha$ and $\beta$ are obtained from the linear
fitting of the isotopic ratio and the isotonic ratio in the chosen reactions
according to Eq. (\ref{isoscaling}). For example, in Fig. \ref{ISCa4048}, the
isoscaling phenomena of the fragments in the 140$A$ MeV $^{40,48}$Ca + $^9$Be
reactions are shown. In Figs. \ref{ISCa4048}(a) and (b), the isotopic scaling
and the isotonic scaling are plotted, respectively (similar results can be
found in Ref. \cite{FuY09isoCaNi}). $\alpha$ ($\beta$) equal to the slope of
the linear fitting of the isotopic (isotonic) scaling. For one fragment, the
IS-$\Delta\mu/T$ is calculated using the $\alpha$ and $\beta$ obtained from its
$Z$ isotopes and $N$ isotones, respectively. For IB-$\Delta\mu/T$, the IYR in
each reaction is calculated first, then the difference between the IYRs in the
two reactions is calculated according to Eq. (\ref{DBratio}).

\begin{figure}[htbp]
\includegraphics
[width=8.cm]{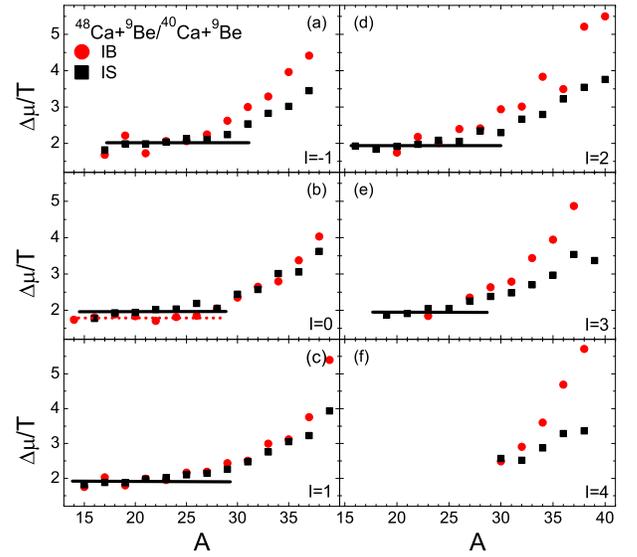}
\caption{\label{Ca40Ca48DR} (Color online) The IB- and
IS-$\Delta\mu/T$ in the 140$A$ MeV $^{40,48}$Ca + $^9$Be
reactions \cite{Mocko06}. $I=N-Z$ is the neutron-excess.
The lines are for guiding the eyes to the plateaus.
}
\end{figure}

In Fig. \ref{Ca40Ca48DR}, the IB- and IS-$\Delta\mu/T$ in the
$^{48}$Ca/$^{40}$Ca + $^9$Be reactions are plotted. Very similar trends of the IB-
and IS-$\Delta\mu/T$ distributions in each $I$-chain of fragments are found. In each $I$-chain,
both the IB- and IS-$\Delta\mu/T$ in the small-$A$ fragments form plateaus
(around $\Delta\mu/T = 2$), and $\Delta\mu/T$ increases as $A$ increases in
the large-$A$ fragments. The plateuas of the IB- and IS-$\Delta\mu/T$ almost
overlap, and it is interesting that the IB- and IS-$\Delta\mu/T$ have very little
difference in the $I=0$ [panel (b)] and the $I=1$ [panel (c)] fragments. Except for
the $I=0$ and $I=1$ fragments, differences between the IB- and IS-$\Delta\mu/T$
in the large-$A$ fragments are shown.

\begin{figure}[htbp]
\includegraphics
[width=8.6cm]{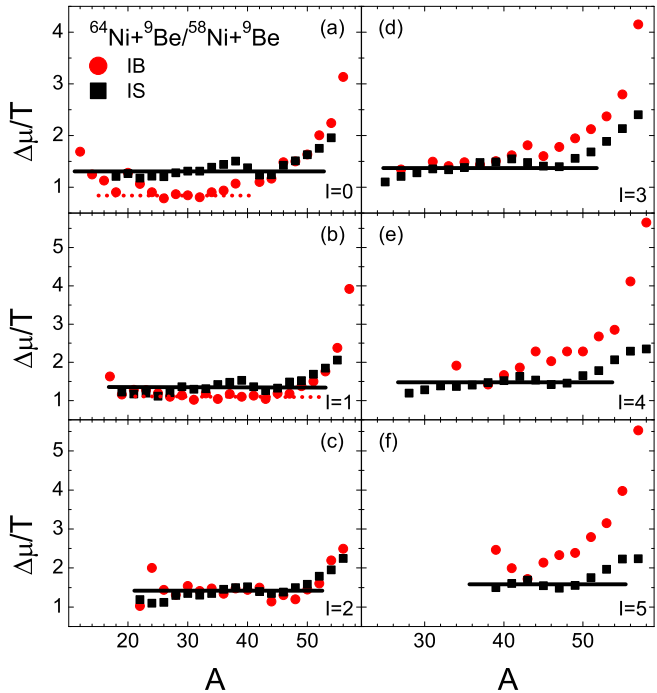}
\caption{\label{Ni58Ni64DR}(Color online)  The IB- and IS-$\Delta\mu/T$ in the
140$A$ MeV $^{58,64}$Ni + $^9$Be reactions \cite{Mocko06}. The lines are
for guiding the eyes to the plateaus.
}
\end{figure}

In Fig. \ref{Ni58Ni64DR}, the IB- and IS-$\Delta\mu/T$ in the $^{64}$Ni/$^{58}$Ni + $^9$Be
reactions are plotted. Very similar results as the $^{48}$Ca/$^{40}$Ca reactions are found,
except that the values of the plateaus decrease to about 1.4. The values of the IB- and
IS-$\Delta\mu/T$ almost overlap in the $I=1$ [panel (b)] and the $I=2$ [panel (c)] fragments.
The $n/p$ for $^{48}$Ca/$^{40}$Ca is 1.4/1.0, which is larger than the value 1.286/1.071 for
$^{64}$Ni/$^{58}$Ni. Comparing to the results in the $^{48}$Ca/$^{40}$Ca reactions, obviously larger
widths of the IB- and IS-$\Delta\mu/T$ plateaus in the $^{64}$Ni/$^{58}$Ni reactions
are found.

\begin{figure}[htbp]
\includegraphics
[width=8.6cm]{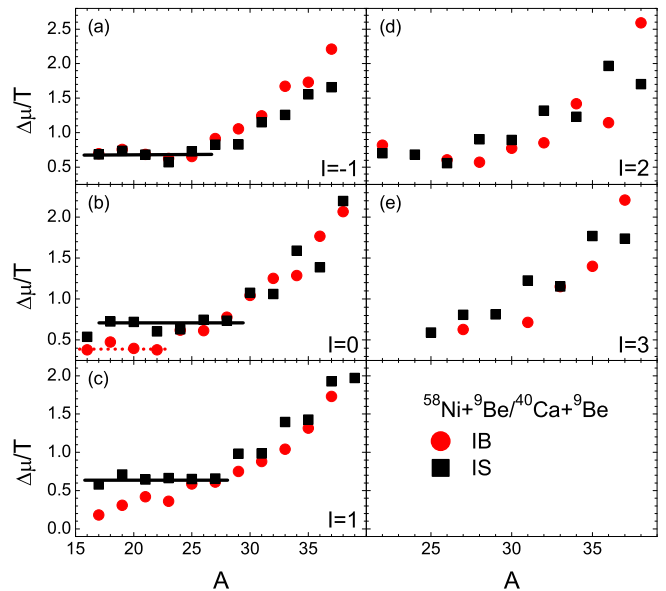}\caption{\label{Ca40Ni58DR} (Color online)
 The IB- and IS-$\Delta\mu/T$ in the 140$A$ MeV
$^{40}$Ca + $^9$Be and $^{58}$Ni + $^9$Be reactions  \cite{Mocko06}.
The lines are just for guiding the eyes of the plateaus.
}
\end{figure}

From the results shown in Figs. \ref{Ca40Ca48DR} and \ref{Ni58Ni64DR}, for the
plateau, its height (the value) and its width (the nuclei range it covers) should
be noticed. In the statistical models \cite{MaCW09PRC,MaCW10PRC}, the yield of a
fragment to some extent is decided by the density distributions of protons ($\rho_p$)
and neutrons ($\rho_n$) of the projectile and the target nuclei. A nucleus can
be assumed to have a core region, in which $\rho_p$ and $\rho_n$ change very
little, and a skirt region, in which $\rho_p$ and $\rho_n$ change fast. The $\rho_p$
distributions of isotopes can be assumed to be similar, especially when the masses
of the isotopes do not differ much. This indicates that, the height
of the plateau is decided by the difference between $\rho_n$ and $\rho_p$ in
the projectiles, and the width shows the overlapping volume of the projectiles
in which $\rho_n$ and $\rho_p$ vary slowly. Regarding the $\Delta\mu/T$ in the
isotopic projectile reactions, the height of the plateau indicates the difference
between the $\rho_n$ distributions of the projectiles.

In addition to the similarity of the isotopic distributions in the $^{40}$Ca/$^{48}$Ca and
the $^{58}$Ni/$^{64}$Ni reactions, the similarity of the isotopic and the isotonic distributions
in the $^{58}$Ni/$^{40}$Ca and the $^{48}$Ca/$^{64}$Ni reactions were also investigated
in Refs. \cite{MaCW09PRC,MaCW09CPB}. It is found that the isotopic or isotonic
yield distribution shows dependence on $\rho_n$ and $\rho_p$ of the projectiles.
In the isoscaling method, the system effects do influence $\alpha$ and $\beta$. In the IBD
method, the system effects are also removed in the IYR. Thus the $\Delta\mu/T$
in the \textit{n/p} symmetric $^{58}$Ni/$^{40}$Ca reactions, and the neutron-rich
$^{48}$Ca/$^{64}$Ni reactions can also be analyzed. In Fig. \ref{Ca40Ni58DR}, the
$\Delta\mu/T$ in the $^{58}$Ni/$^{40}$Ca reactions are plotted. The \textit{n/p}
of $^{58}$Ni/$^{40}$Ca is 1.071/1.0, thus $\rho_n$ and $\rho_p$ for them can be assumed
to have similar trends but differ in values. In Fig. \ref{Ca40Ni58DR}, very small
differences between the IB- and IS-$\Delta\mu/T$ are found in each $I$-chain.
The values of the plateaus decrease to about 0.7, and in the $I = $0 and 1 chain,
 the IB-plateaus become even smaller.

\begin{figure}[htbp]
\includegraphics
[width=8.6cm]{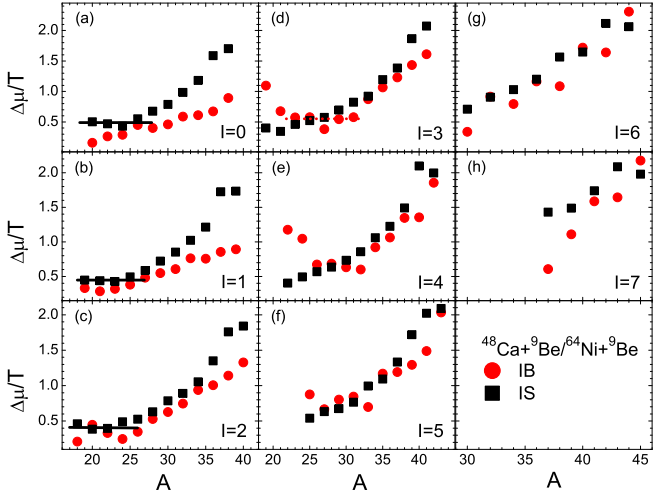}\caption{\label{Ca48Ni64DR} (Color online)
 The IB- and IS-$\Delta\mu/T$ in the
140$A$ MeV $^{48}$Ca + $^9$Be and $^{64}$Ni + $^9$Be
reactions \cite{Mocko06}. The lines are for guiding
the eyes to the plateaus.
}
\end{figure}

The IB- and IS-$\Delta\mu/T$ in the $^{48}$Ca/$^{64}$Ni reactions
are plotted in Fig. \ref{Ca48Ni64DR}. The \textit{n/p} ratio of $^{48}$Ca/$^{64}$Ni
is 1.4/1.286. Though large differences between the IB- and IS-$\Delta\mu/T$ in the
$I\leq2$ fragments are shown, similar values of the IB- and IS-$\Delta\mu/T$ in
the $I\geq3$ ($A\geq25$) fragments are found. The plateaus in the $I\leq3$ chains
decrease to smaller than 0.5. The characteristics of the IB- and IS-$\Delta\mu/T$
distributions are very similar to those of the isotopic or isotonic distributions
shown in Ref. \cite{MaCW09PRC}, i.e., the density dependence of $\rho_n$ and $\rho_p$
in the projectiles.

In Fig. \ref{Ca40Ca48DR}-\ref{Ca48Ni64DR}, similarities between IB- and IS-$\Delta\mu/T$ are shown,
i.e., the plateaus and the values in the small-$A$ fragments of the $^{48}$Ca/$^{40}$Ca
and $^{64}$Ni/$^{58}$Ni reactions, most fragments of the $^{58}$Ni/$^{40}$Ca, and large-$A$ fragments of the $^{48}$Ca/$^{64}$Ni reactions; while the differences between
the IB- and IS-$\Delta\mu/T$ are shown in the large-$A$ fragments of the $^{48}$Ca/$^{40}$Ca
and $^{64}$Ni/$^{58}$Ni reactions. The values of the plateaus show a dependence of
the \textit{n/p} ratio of the reaction systems, which are about 2, 1.4, 0.7, and 0.5 in
the $^{48}$Ca/$^{40}$Ca, $^{64}$Ni/$^{58}$Ni, $^{58}$Ni/$^{40}$Ca, and $^{48}$Ca/$^{64}$Ni
reactions.

\begin{figure}[htbp]
\includegraphics
[width=8.6cm]{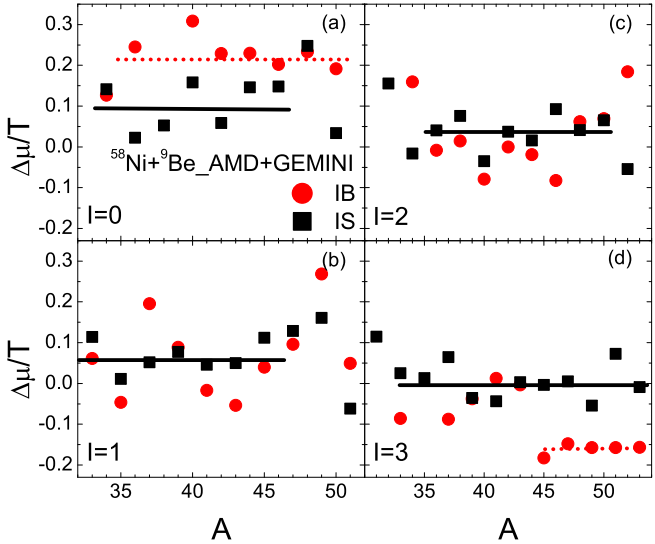}\caption{\label{Ni58CentPeriAMD} (Color online)
 The IB- and IS-$\Delta\mu/T$ between the fragments
in the impact parameter regions of R1 (\textit{b}=0-2 \textit{fm}) and R2 (\textit{b}=6-8 \textit{fm})
in the 140$A$ MeV $^{58}$Ni + $^9$Be reactions calculated using
the AMD + GEMINI models.
}
\end{figure}

To see more clearly on this point, we performed a simulation of the 140$A$ MeV
$^{58}$Ni + $^{9}$Be reaction using a microscopic transport model. Though there
are many choices-- such as the quantum molecular dynamics (QMD and its improved
versions) models \cite{ZhouPei11T,IQMDmodel,ImQMD,TianWD05-iso-CPL,SuJun11PRCTiso,Kumar11-12PRC-iqmd},
the AMD model \cite{Ono03RRCiso,Ono04RRCiso,OnoAMD96,OnoAMD99}, and different
methods to form the cluster or fragment \cite{iqmdbind,GEMINI}-- the AMD model
plus the sequential decay GEMINI code \cite{GEMINI} have been used to simulate the
reaction, since similar works are performed, and the experimental yields of
fragments are well reproduced \cite{Mocko08-AMD}. In the calculation, the
standard Gogny interaction (Gogny-g0) is used \cite{Gogny-g0}, the fragments
are formed using a coalescence radius, $R_c = 5fm$ in the phase space at the time
$t = 500 fm/c$ in AMD. Two cuts of impact parameters are used in the fragments
analysis, i.e., \textit{b}=0-2 \textit{fm} (labeled as R1) and \textit{b}=6-8\textit{fm}
(labeled as R2). The results are plotted in Fig. \ref{Ni58CentPeriAMD}. Since
$^{58}$Ni is a symmetric nucleus, its $\rho_n$ and $\rho_p$ distributions are
very similar, and the densities only decrease sharply in the very edge of the
nucleus. In R1 and R2, the difference between $\rho_n$ and $\rho_p$ is also
very small. According to the assumption that the plateau of $\Delta\mu/T$ depends
on $n/p$ of the projectiles, the plateaus of the IB-$\Delta\mu/T$ should be
very small. In Fig. \ref{Ni58CentPeriAMD}, it can be seen that the IB- and
IS-$\Delta\mu/T$ are very similar, and their values are very small. The
IS-$\Delta\mu/T$ of most fragments are in a range of $0.1\pm0.1$, and the
IB-$\Delta\mu/T$ of the $I=0-2$ chains are $0\pm0.1$. This indicates that the
IB- and the IS-$\Delta\mu/T$ have very little difference when $\rho_n$ and
$\rho_p$ of the two projectiles are similar, and differ if $\rho_n$ and $\rho_p$
of the two projectiles are different.

Furthermore, the difference between the IB- and IS-$\Delta\mu/T$ should also be
discussed. Generally, the IB- and IS-$\Delta\mu/T$ should be the same since they
are obtained from the same fragments and in the same theory. Though in most
fragments the IB- and IS-$\Delta\mu/T$ are very similar, the difference between
the IB- and IS-$\Delta\mu/T$ are also shown in many fragments. For examples,
in fragments where the IB- and IS-$\Delta\mu/T$ are different, the IB-$\Delta\mu/T$
are larger than the IS-$\Delta\mu/T$ in the $^{48}$Ca/$^{40}$Ca and $^{64}$Ni/$^{58}$Ni
reactions, while the IB-$\Delta\mu/T$ are smaller than the IS-$\Delta\mu/T$ in
the $^{58}$Ni/$^{40}$Ca and $^{48}$Ca/$^{64}$Ni reactions. According to Eq.
(\ref{isoscaling}), the isoscaling parameters $\alpha$ ($\beta$) are obtained
from the linear correlation between the isotopic (isotonic) ratio and neutron
(proton) numbers, and $\alpha$ ($\beta$) is the same for all the isotopes (isotones).
In other words, $\alpha$ ($\beta$) is the scaled parameter for all the isotopes
(isotones). For one fragment, its $\alpha$ ($\beta$) is constrained by its isotopes
(isotones), thus $\alpha$ ($\beta$) can not reflect the difference between the
isotopes or isotones. While according to Eq. (\ref{DBratio}), the IB-$\Delta\mu/T$
result only relies on the two related isobars, and the difference between isobars
of different masses can be obviously shown. It is suggested that since the IBD method
uses only two isobars, the IB-$\Delta\mu/T$ result is not influenced by the rest
fragments, and more precise results than the isoscaling method can be obtained, especially
when the fragment disobeys the isoscaling.

Finally, we discuss the temperature effect in the $\Delta\mu/T$. $\mu_n$ and
$\mu_p$ depend on both the density and the temperature. In theories based on the free energy,
it is difficult to separate the free energy and the temperature \cite{Huang10,MaCW11PRC06IYR}.
In the IB- and the IS-$\Delta\mu/T$, it is also difficult to separate $\Delta\mu$
and $T$. Besides the density effects in $\Delta\mu/T$, the temperature also influences
$\Delta\mu/T$. Actually, the temperature should be defined at thermal equilibrium,
but in intermediate energy HICs no thermal equilibrium is reached. In other words,
the temperature is nonuniform in the collisions. In QMD, the "temperature" can be
extended to the non-equilibrium situations and extracted in the local density
approximation \cite{PuriNPA94}. In grand-canonical ensemble theory, the temperature
is supposed to be the same, but differs in each reaction system. In a recent work
using a canonical thermodynamic model, a temperature profile of impact parameter
($b$) is introduced, in which the temperature decreases as $b$ increases
\cite{Mallik11PRC-PF-Tb}. Considering the multiple sources collisions of different
strengths according to the impact parameters, the temperature changes with the excitation
energy. In the Fermi-gas relationship the correlation between the excitation energy
per nucleon ($E^{*}/A$) and temperature is $E^{*}/A = T^2/\mbox{a}$, or $T = \sqrt{E^{*}a/A}$,
in which $\mbox{a} = Ak$ and $k$ is the inverse level density parameter. In Ref. \cite{FuY09isoCaNi},
in the $^{48}$Ca/$^{40}$Ca + $^9$Be reactions, $\alpha$ is found to decreases when
$E^{*}/A$ increases (which corresponds to $T = 1.2 - 2.14$ MeV), but $\alpha$
tends to be similar if $E^{*}/A$ is high. Noting that $\alpha\approx-\beta$, the
temperature dependence of $\alpha$ can also explain the plateau plus the increasing
part of the IB- and IS-$\Delta\mu/T$ distributions as follows: if $\Delta\mu$ are
uniform in the source, the plateau forms in the fragments which have high $E^{*}/A$,
while the $\Delta\mu/T$ increases when $E^{*}/A$ decreases in the fragments which
have low $E^{*}/A$. In the statistical abrasion-ablation (SAA) model, the excitation
energy is $E^{*}=13.3\Delta A$ MeV, in which $\Delta A$ is the number of nucleons
removed from the projectile by the ablation-abrasion process \cite{MaCW09PRC,MaCW12PRCT}.
Then $T = \sqrt{13.3k\Delta A/A}$. In peripheral collisions, due to the low density
of nucleons, the abraded nucleons are fewer than those in the central collisions, which
results in the relative low temperature. To some extent, the low temperature is a
result of the low density in the peripheral collisions, which is similar to the temperature
profile in Ref. \cite{Mallik11PRC-PF-Tb}. Thus low density can also result in an
increase of $\Delta\mu/T$. In an isobaric method, the temperature of the measured
heavy fragment is suggested to be similar due to the significant influence of the
secondary decay process \cite{MaCW12PRCT}. Thus, though the density dependence and
the temperature dependence of $\Delta\mu/T$ cannot be totally separated, the density
dependence is preferred since the low temperature is one result of the low density.

\section{summary}
\label{summary}
To summarize, in the article, a new IBD method is proposed to investigate $\Delta\mu/T$
of the colliding sources, and the result is compared with the result of the usually used isoscaling
method within the same grand-canonical model. The IB- and IS-$\Delta\mu/T$ are found to
be similar in the distributions, which both have a plateaus in small mass fragments plus
an increasing part in relatively larger mass fragments. The IB- and IS-$\Delta\mu/T$ plateaus
show dependence on the $n/p$ ratio of the projectiles. It is suggested that the height of
the plateau is decided by the difference between $\rho_n$ and $\rho_p$ in the projectiles,
and the width shows the overlapping volume of the projectiles in which $\rho_n$ and $\rho_p$
change very little. The difference between the IB- and IS-$\Delta\mu/T$ is explained
by $\alpha$ and $\beta$ being constrained by the many isotopes and isotones, while the IBD
method only uses the yields of two isobars. It is suggested that the IB-$\Delta\mu/T$ is more
reasonable than the IS-$\Delta\mu/T$, especially when the isotopic or isotonic ratio disobeys
the scaling. As to the question whether $\Delta\mu/T$ depends on the density or the temperature
of the colliding, the density dependence is preferred since the low density can result in the
low temperature in the peripheral reactions.

\begin{acknowledgments}
This work is supported by the National Natural Science Foundation of China under Contract
No. 10905017, the Program for Science\&Technology Innovation Talents in Universities of
Henan Province (HASTIT), and the Young Teacher Project in Henan Normal University (HNU),
China. We also thank the helpful guides of Dr. HUANG Mei-Rong at the Institute of Modern
Physics (IMP), Chinese Academy of Sciences, for the AMD simulation and data analysis.
A useful discussion with Professor Roy Wada at IMP is also acknowledged. The AMD simulation
is performed on the high-performance computing center at the College of Physics and
Electrical Engineering, HNU.
\end{acknowledgments}


\begin{thebibliography}{}
%
%

%

\bibitem{AlbNCA85DRT} 
S. Albergo 
{\it et al.},
Nuovo Cimento A {\bf 89}, 1 (1985).
\bibitem{Isoscaling} Y. G. Ma \textit{et al.}, Phys. Rev. C \textbf{69}, 064610 (2004); Phys. Rev. C \textbf{72}, 064603 (2005).
\bibitem{HShanPRL} H. S. Xu \textit{et al.}, Phys. Rev. Lett. \textbf{85}, 716 (2000).
\bibitem{Bot02-iso-T}
A. S. Botvina 
\textit{et al},
Phys. Rev. C \textbf{65}, 044610 (2002).
\bibitem{MBTsPRL01iso}
M. B. Tsang 
\textit{et al.},
Phys. Rev. Lett. \textbf{86}, 5023 (2001).
\bibitem{Huang10} 
M. Huang 
{\it et al.},
Phys. Rev. C \textbf{81}, 044620 (2010).

\bibitem{ModelFisher3}
A. S. Hirsch 
{\it et al.},
Phys. Rev. C \textbf{29}, 508 (1984).

\bibitem{MaCW12PRCT}
C. W. Ma 
{\it et al.},
Phys. Rev. C \textbf{86}, 054611 (2012).
\bibitem{IS-fluctuation13} M. Colonna, Phys. Rev. Lett. \textbf{110}, 042701 (2013).
\bibitem{BALi08PR}
B.-A. Li 
{\it et al.}, Phys. Rep. \textbf{464}, 113 (2008).
\bibitem{Huang10NPA-Mscaling}
M. Huang 
{\it et al.},
Nucl. Phys. A \textbf{847}, 233 (2011). 
\bibitem{Fang07-iso-JPG}
D. Q. Fang 
{\it et al.},
J. Phys. G: Nucl. Part. Phys. \textbf{34}, 2173 (2007).
\bibitem{FuY09isoCaNi}
Y. Fu 
{\it et al.},
Chin. Phys. Lett. \textbf{26}, 082503 (2009).


\bibitem{ChenZQ10-iso-sym} Z. Chen 
\textit{et al.},
Phys. Rev. C \textbf{81}, 064613 (2010).
\bibitem{Ono03RRCiso}
A. Ono 
\textit{et al.},
Phys. Rev. C \textbf{68}, 051601(R) (2003).
\bibitem{Ono04RRCiso}
A. Ono 
\textit{et al.},
Phys. Rev. C \textbf{70}, 041604(R) (2004).

\bibitem{SouzaPRC09isot} 
S. R. Souza 
\textit{et al.},
Phys. Rev. C \textbf{80}, 044606 (2009).
%
\bibitem{Soul03-iso}
G. A. Souliotis 
\textit{et al.},
Phys. Rev. C \textbf{68}, 024605 (2003).
\bibitem{Soul06-iso-T-sym}
G. A. Souliotis 
\textit{et al.},
Phys. Rev. C \textbf{73}, 024606 (2006).
\bibitem{Dorso06-iso-finite}
C. O. Dorso, Phys. Rev. C \textbf{73}, 034605 (2006).
\bibitem{TianWD05-iso-CPL}
W. D. Tian 
\textit{et al.},
Chin. Phys. Lett. \textbf{22}, 306 (2005).


\bibitem{ZhouPei11T} 
P. Zhou 
{\it et al.},
Phys. Rev. C \textbf{84}, 037605 (2011).

\bibitem{Tsang01-iso-sym} 
M. B. Tsang 
{\it et al.},
Phys. Rev. C \textbf{64}, 054615 (2001).
\bibitem{Liu04-iso-sym} 
T. X. Liu 
{\it et al.},
Phys. Rev. C \textbf{69}, 014603 (2004).
\bibitem{Tsang06-iso}
M. Colonna 
{\it et al.},
Eur. Phys. J. A \textbf{30}, 165 (2006).




\bibitem{MA12CPL09IYRAsbsbv}
C. W. Ma 
{\it et al.},
Chin. Phys. Lett. \textbf{29}, 092101 (2012).

\bibitem{MaCW11PRC06IYR}
C. W. Ma 
{\it et al.},
Phys. Rev. C \textbf{83}, 064620 (2011).

\bibitem{MaCW12EPJA}
Chun-Wang Ma 
{\it et al.},
Eur. Phys. J. A \textbf{48}, 78 (2012). 

\bibitem{MaCW12CPL06}
C.-W. Ma 
{\it et al.},
Chin. Phys. Lett. \textbf{29}, 062101 (2012); Chin. Phys. C \textbf{37}, 024102 (2013).


\bibitem{Goodman84Tc}
A. L. Goodman 
{\it et al.},
Phys. Rev. C \textbf{30}, 851 (1984).

\bibitem{YgMa99PRCTc}
Y. G. Ma 
{\it et al.},
Phys. Rev. C \textbf{60}, 024607 (1999).
\bibitem{YgMa05PRCTc}
Y. G. Ma 
{\it et al.},
Phys. Rev. C \textbf{71}, 054606 (2005).
\bibitem{Nato95T}
J. B. Natowitz 
{\it et al.},
Phys. Rev. C \textbf{52}, R2322 (1995).
\bibitem{SuJun11PRCTiso}
J. Su and F. S. Zhang, Phys. Rev. C {\bf 84}, 037601 (2011);
J. Su, {\it et al.},
Phys. Rev. C \textbf{85}, 017604 (2012).

\bibitem{JSWangT05}
J. Wang 
{\it et al.},
Phys. Rev. C \textbf{72}, 024603 (2005).

\bibitem{Wada97T}
R. Wada 
{\it et al.},
Phys. Rev. C \textbf{55}, 227 (1997).
\bibitem{TrautTHFrg07}
W. Trautmann \textit{et al.} (ALADIN Collaboration),
Phys. Rev. C {\bf 76}, 064606 (2007).
\bibitem{MaCW13ComTheoPhys} 
C. W. Ma, \textit{et al.},
Commun. Theo. Phys. \textbf{59}, 95 (2013).


\bibitem{GrandCan} 
C. B. Das 
{\it et al.},
Phys. Rev. C \textbf{64}, 044608 (2001).

\bibitem{Tsang07BET} 
M. B. Tsang 
\textit{et al.},
Phys. Rev. C \textbf{76}, 041302(R) (2007).
\bibitem{Huang10Powerlaw}
M. Huang 
\textit{et al.},
Phys. Rev. C \textbf{82}, 054602(R) (2010).
\bibitem{Mocko06}
M. Mocko 
{\it et al.},
Phys. Rev. C \textbf{74}, 054612 (2006).

\bibitem{MaCW09PRC}
C. W. Ma 
{\it et al.},
Phys. Rev. C \textbf{79}, 034606 (2009).

\bibitem{MaCW10PRC}
C.-W. Ma 
{\it et al.},
Phys. Rev. C {\bf 82}, 057602 (2010).
\bibitem{MaCW09CPB} C. W. Ma 
{\it et al.},
Chin. Phys. B \textbf{18}, 4781 (2009).
\bibitem{Mocko08-AMD}
M. Mocko 
{\it et al.},
Phys. Rev. C \textbf{78}, 024612 (2008).

\bibitem{IQMDmodel} S. Kumar \textit{et al.}, Phys. Rev. C \textbf{81},014611 (2010);
S. Gautam \textit{et al.}, J. Phys. G: Nucl. Part. Phys. \textbf{37},085102 (2010);
S. Gautam \textit{et al.}, Phys. Rev. C \textbf{83},014603 (2011); \textit{ibid.},
Phys. Rev. C \textbf{83}, 034606 (2011).
\bibitem{ImQMD} N. Wang, Z. Li, and Z. Wu, Phys. Rev. C \textbf{65}, 064608 (2002); Y. Zhang and Z. Li, Phys. Rev. C \textbf{71}, 024604 (2005);
\textit{ibid.}, Phys. Rev. C \textbf{74}, 014602 (2006);
Y. Zhang 
\textit{et al.,}
Phys. Lett. B \textbf{664}, 145 (2008).
\bibitem{Kumar11-12PRC-iqmd}
S. Kumar 
\textit{et al.,}
Phys. Rev. C \textbf{84}, 044620 (2011);
S. Kumar 
\textit{et al.,}
Phys. Rev. C \textbf{86}, 044616 (2012).
\bibitem{iqmdbind}
S. Goyal and R. K. Puri, Phys. Rev. C. \textbf{83}, 047601 (2011);
Y. K. Vermani \textit{et al.}, J. Phys. G: Nucl. Part. Phys. \textbf{37}, 015105 (2010);
R. K. Puri, and J. Aichelin, J. Comp. Phys. \textbf{162}, 245(2000). 
\bibitem{OnoAMD96}
A. Ono and H. Horiuchi, Phys. Rev. C \textbf{53}, 2958 (1996).
\bibitem{OnoAMD99}
A. Ono, Phys. Rev. C \textbf{59}, 853 (1999).
\bibitem{GEMINI}
R. J. Charity 
\textit{et al.,} Nucl. Phys. A \textbf{483}, 371 (1988).
\bibitem{Gogny-g0} J. Decharg\'{e} and D. Gogny, Phys. Rev. C \textbf{21}, 1568 (1980).

\bibitem{PuriNPA94}
R. K. Puri 
\textit{et al.,} Nucl. Phys. A \textbf{575}, 733 (1994).

\bibitem{Mallik11PRC-PF-Tb}
S. Mallik 
\textit{et al.,}
Phys. Rev. C \textbf{84}, 054612 (2011).

%
%
%
%
%
%
%
%
%
%
%







%
%



%
%
%

%
%


%







%


%
%






%

%


\end{thebibliography}
\end{document}